\documentclass[sigconf, nonacm]{acmart}
\AtBeginDocument{%
  }

\setcopyright{acmlicensed}
\copyrightyear{2018}
\acmYear{2018}
\acmDOI{XXXXXXX.XXXXXXX}

\acmConference[Conference acronym 'XX]{Make sure to enter the correct
  conference title from your rights confirmation emai}{June 03--05,
  2018}{Woodstock, NY}
\acmISBN{978-1-4503-XXXX-X/18/06}

\usepackage{amsmath,amsfonts}
\usepackage{algorithmic}
\usepackage{graphicx}
\usepackage{textcomp}
\usepackage{xcolor}
\usepackage{caption}
\usepackage{tikz}
\usepackage{multirow}
\usepackage{float}
\usepackage{subcaption} 
\usepackage{booktabs}
\usepackage{algorithm}
\usepackage{array}

\begin{document}


\title{Towards Robust Real-Time Hardware-based Mobile Malware Detection using Multiple Instance Learning Formulation
}

\author{Harshit Kumar, Sudarshan Sharma, Biswadeep Chakraborty, Saibal Mukhopadhyay}
\affiliation{%
  \institution{Electrical and Computer Engineering, Georgia Instititute of Technology}
  \city{Atlanta}
  \state{Georgia}
  \country{USA}\\
{\footnotesize \color{red}Under peer review.}
}

\renewcommand{\shortauthors}{Kumar et al.}

\begin{abstract}
This study introduces RT-HMD, a Hardware-based Malware Detector (HMD) for mobile devices, that refines malware representation in segmented time-series through a Multiple Instance Learning (MIL) approach. We address the mislabeling issue in real-time HMDs, where benign segments in malware time-series incorrectly inherit malware labels, leading to increased false positives. Utilizing the proposed Malicious Discriminative Score within the MIL framework, RT-HMD effectively identifies localized malware behaviors, thereby improving the predictive accuracy. Empirical analysis, using a hardware telemetry dataset collected from a mobile platform across 723 benign and 1033 malware samples, shows a 5\% precision boost while maintaining recall, outperforming baselines affected by mislabeled benign segments.
\end{abstract}



\maketitle

\section{Introduction}
 Recent years have seen a surge in malicious applications challenging the security of modern devices \cite{fireeye}. This has shifted Endpoint Security towards \textit{behavior analysis}, using sensors across the compute stack for \textit{continuous} monitoring and data analysis \cite{microsoft_endpoint}. To address the limitations of high performance overhead of traditional software-based approaches, HMD has emerged, utilizing low-level hardware telemetry as a behavioral signature for low-overhead and tamper-resistant detection of malicious activities \cite{demme_hpc, intel_tdt}. HMD, a vital component in collaborative defense systems such as Endpoint Detection and Response (EDR) \cite{intel_tdt, noauthor_qualcomm_2017}, enhances threat visibility and complements software-level techniques, effectively expanding the scope of detected threats \cite{quant_hmd,kumar_xmd}. 

HMDs offer computational efficiency through GPU offloading \cite{intel_tdt} and hardware acceleration \cite{map}, allowing for real-time malware monitoring with minimal system impact. Central to HMDs is a Machine Learning (ML) classifier that performs binary classification on segmented time-series data for real-time detection. Selecting the optimal window length in HMDs is crucial, as it balances the need for capturing comprehensive malware behavior for effective training \cite{sandboxExecution} with the necessity for early detection during inference~\cite{map}—\textit{longer windows improve learning while shorter ones enable quicker response times}. To balance these conflicting requirements, HMDs typically segment long-duration time-series into shorter windows for ML training, with window labels inherited from the parent time-series \cite{map, ensemblehmd,basu_tsc_malware_detection, low_barrier_hmd}. \textbf{We hypothesize that the direct inheritance of labels from time-series to windows inaccurately represents malware due to prevalent benign segments within malware-labeled time-series.} Such labeling inaccuracies, combined with the high base rate of benign workloads in malware detection \cite{ml_pitfall_arp}, contribute to an increase in false positives, leading to alarm fatigue among security analysts \cite{false_positive}.

Drawing on Demme et al.'s discussion of malware labeling challenges \cite{demme_hpc}, we evaluate the common practice of applying a single malware label to all the windows of a segmented time-series. A malware-infected time-series contains both malicious and benign segments, yet conventional labeling does not distinguish between them. This labeling approach overlooks the localized nature of malware behaviors, which are typically restricted to certain time-series segments, evidenced by key malicious code snippets in software binaries (e.g. Command-and-Control communication, File Discovery, etc.). Consequently, benign-only segments are mis-labeled as malicious for training a ML Classifier. To mitigate this, we propose a Multiple Instance Learning (MIL) formulation \cite{foulds_frank_2010_mil} that accurately reflects the localized malware behaviors, thereby improving the alignment of malware representation with the actual distribution of malicious activity within the time-series, and improving the performance of the ML Classifier.

\begin{figure}[!bp]
    \centering
    \includegraphics[width = 0.47\textwidth]{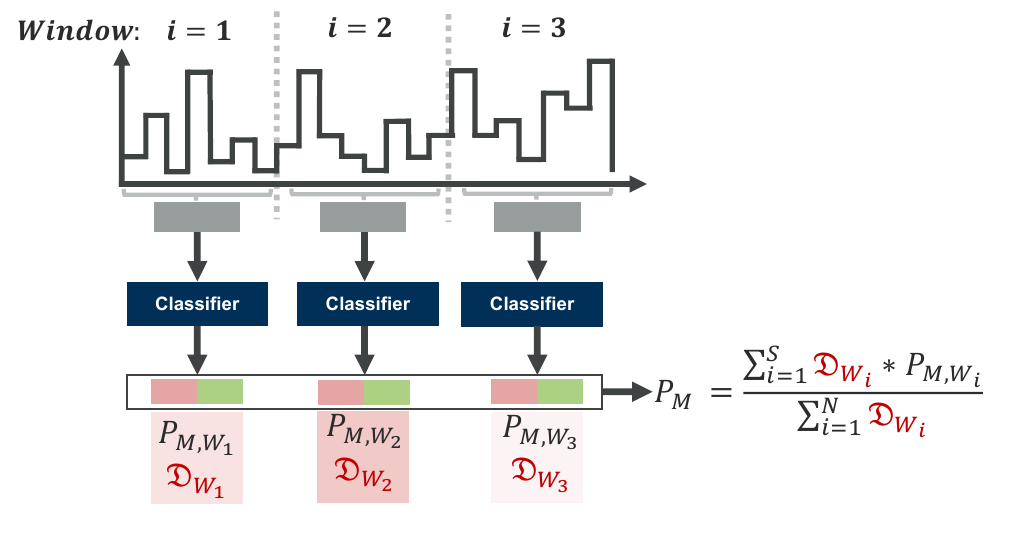}
    \caption{Proposed Methodology: The decision for i-th window \(P_{M,W_i}\) is enhanced by the Malicious Discriminative Score \(\mathfrak{D}_{W_i}\). This score adjustment corrects mis-classifications of benign segments within a malware-labeled time-series, thereby improving the precision of malware detection.}
    \label{fig:main_fig}
\end{figure}

Our empirical study leverages the HMD model dataset from Kumar et al. \cite{kumar_xmd}, examining multivariate time-series data derived from SoC components using Dynamic Voltage and Frequency Scaling (DVFS) and SYSFS telemetry.\textit{ We introduce the Malicious Discriminative Score (MDS) to evaluate the distinctivenes of segment behaviors, assigning higher scores to segment-behaviors uniquely observed in malware time-series.} This score enhances the classifier's output by amplifying the signal for distinct malware behavior and attenuating it for benign behavior, thus fine-tuning the prediction for each time segment. Illustrated in Figure \ref{fig:main_fig}, the prediction for each window ($P_{M,W_i}$) is adjusted by its corresponding MDS ($\mathfrak{D}_{W_i}$), and the overall prediction $P_{M}$ is computed as the weighted sum of these adjusted segment predictions. To calculate the MDS, we model first-order interactions between time-series channels using conditional histograms, learning complex behavioral patterns as compared to single channel analysis which treat each channel of the multivariate time-series independently. Our key contributions are:
\begin{itemize}
    \item Introduction of MIL formulation for accurate representation of malware behavior in segmented time-series, to reduce False Positives in real-time HMDs.
    \item Proposal of MDS to support the MIL assumption, calculated using a novel statistical classifier by analyzing first-order interactions in multivariate time-series.
\end{itemize}
The rest of the paper is organized as follows. We provide a brief background in Section \ref{sec:background}. Next, we explain our proposed framework in Section \ref{sec:proposed_methodology}. We present our evaluation results and discussion in Section \ref{sec:eval_results} and conclude in Section \ref{sec:conclusion}.

\section{Background}
\label{sec:background}

\subsection{HMD and Related Works}
Demme et al. leveraged Hardware Performance Counters (HPCs) to detect a range of malicious activities \cite{demme_hpc}. Subsequent research has focused on improving HMDs' predictive capabilities \cite{ensemblehmd, kumar_uncertainty} and reducing their performance overhead \cite{analyze_hmd, low_barrier_hmd, map}. Alternatives to HPCs, such as DVFS telemetry \cite{chawla_iotj}, have also been explored for their system-wide behavior profiling across various SoC sub-devices, widening the scope of detectable attacks beyond those with prominent architectural side-channel effects \cite{kumar_xmd}. 

Efforts to address real-time detection challenges have been noted \cite{ratfia, map, ensemblehmd,basu_tsc_malware_detection, low_barrier_hmd}. Alam et al. used an anomaly detection approach based on reconstruction error for real-time detection, avoiding direct benign vs. malware classification and focusing on normal vs. anomalous behavior \cite{ratfia}. However, the variability in defining "normal" often leads to high false positives in anomaly detection. In contrast, the benign vs. malware classification paradigm persists \cite{map, ensemblehmd,basu_tsc_malware_detection, low_barrier_hmd}. While a broad classification of the entire time-series as malware might be justifiable (given the presence of malicious code) as done in prior works \cite{demme_hpc}, using the parent labels for the windows is inadequate as it mislabels benign behavior. Kuruvila et al. suggested using ensemble of classifiers operating on extended window lengths to reduce false positives \cite{basu_tsc_malware_detection}. However, increased window lengths delay early detection. Our method diverges by preserving the window length, enabling timely detection, and utilizing the uniqueness of malicious behavior to fine-tune the window predictions, reducing false positives.

\subsection{Multiple Instance Learning Formulation and Its Implementation Challenges}
The crux of MIL is distinguishing between `bags'—the entire time-series—and `instances'—the segmented windows within \cite{foulds_frank_2010_mil}. Our MIL assumption assumes that a bag acquires a malware label if any instance within exhibits malware characteristics, reflecting the locality in malicious behavior within a malware time-series. This deviates from the current assumption where all instances inherit the bag's label \cite{map, ensemblehmd,basu_tsc_malware_detection, low_barrier_hmd}. A primary challenge in MIL is the absence of segment-level labels, which would otherwise require intensive forensic analysis to assign accurately. Such granularity in labeling is both impractical and unscalable due to the resources it demands. We address this by adopting a data-driven approach to learn what constitutes a malicious segment within a time-series: \textit{it must be unique to malware-labeled bags within the training dataset, and exhibit significant deviation from its benign counterpart}. This operational definition aims to isolate high-discriminatory  malicious segments, enhancing the fidelity of malware detection. 

\begin{table}[!tbp]
\centering
\caption{Hardware telemetry channels in the dataset \cite{kumar_xmd}}
\label{tab:dvfs_sysfs_channels}
\resizebox{0.44\textwidth}{!}{%
\begin{tabular}{cll}
\toprule
\textbf{Telemetry Class} & \textbf{Channel \#} & \textbf{Description} \\
\midrule
\multirow{11}{*}{\textbf{DVFS}} & 1 & GPU controller \\
 & 2 & CPU controller: lower cluster \\
 & 3 & CPU controller: higher cluster \\
 & 4 & CPU bus bandwidth controller \\
 & 5 & GPU bus bandwidth controller \\
 & 6 & GPU bus bandwidth controller \\
 & 7 & Latency controller L3 cache: lower cluster \\
 & 8 & Latency controller L3 cache: higher cluster \\
 & 9 & Last level cache controller \\
 & 10 & Memory latency controller: lower cluster \\
 & 11 & Memory latency controller: higher cluster \\
\midrule
\multirow{4}{*}{\textbf{SYSFS}} & 12 & Network: received bytes \\
 & 13 & Network: transmitted bytes \\
 & 14 & Device current \\
 & 15 & Device voltage \\
\bottomrule
\end{tabular}%
}
\end{table}

\subsection{Dataset Information}
We use the dataset from Kumar et al. \cite{kumar_xmd}, consisting of telemetry data focusing on DVFS and SYSFS states in an SoC. The telemetry is collected on a commodity Android smartphone. The DVFS states, crucial for power management, reflect the workload-induced activity in CPU, GPU, buses, caches, memory, and network interfaces (see Table \ref{tab:dvfs_sysfs_channels}). The dataset contains 1033 malware and 723 benign Android applications sourced from AndroZoo between December 2019 and June 2021. The malware apps span 54 families, verified by the ESET-NOD32 AV engine on VirusTotal. Each app's hardware telemetry signature, collected over eight iterations, results in 2,120 signatures for malicious and 2,143 for benign apps, with each signature being a multivariate-time-series capturing 40 seconds of application execution. We employ a 70-30 train-test split with the test set comprising applications absent from the training set.

\subsection{Threat Model}
RT-HMD assumes attackers can deploy malware to manipulate application-level activities without root access, similar to \cite{kumar_xmd}. Designed for multi-core mobile devices, RT-HMD operates under the premise that users engage with a limited number of foreground applications (1-2) simultaneously. This assumption is due to the reliance on system-wide telemetry channels (Table \ref{tab:dvfs_sysfs_channels}), capturing data from all device processes. RT-HMD's training necessitates malware behavior profiling within a controlled sandbox environment. Consequently, any sandbox evasion techniques could compromise this profiling process, impacting the quality of training data and, ultimately, malware detection effectiveness in real-world scenarios. Additionally, RT-HMD is vulnerable to collusion-based attacks, where attacker-controlled threads could interfere with system-wide telemetry, potentially skewing detection capabilities.

\begin{figure*}[!ht]
    \centering
    \includegraphics[width = 0.80\textwidth]{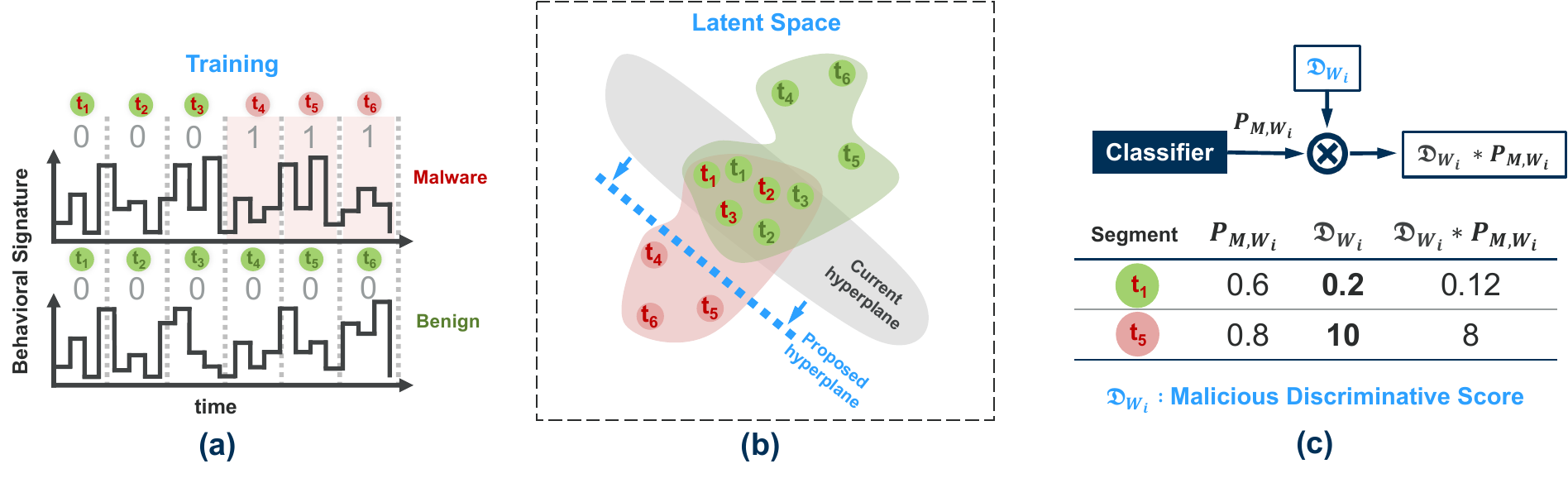}
    \caption{Overview of the Malicious Discriminative Score (MDS): (a) [Top] Malware time-series with annotated benign and malicious segments [Bottom] Segmented benign time-series; (b) hyperplane under strong supervision assumption (current) and hyperplane under MIL assumption (proposed); (c) MDS functionality explained.}
    \label{fig:mds}
\end{figure*}

\section{Proposed Framework}
\label{sec:proposed_methodology}

\subsection{Overview}
The key problem in malware detection using segmented time-series data is the presence of benign segments in both malware and benign-labelled series, leading to conflicting labels and noisy supervision during model training. Figure \ref{fig:mds}.a illustrates this, depicting a malware (1) and a benign (0) time-series, each segmented and labeled. Notably, we can observe benign segments within the malware time-series. These segments receive conflicting labels, blurring the classification boundary. In Figure \ref{fig:mds}.b, we visualize the classifier's latent space, partitioned into three regions: segments unique to malware (lower left), common benign segments (middle), and segments unique to benign time-series (upper right). The overlap of benign segments in both malware and benign time-series complicates the classifier's decision-making, resulting in high false positives.

To address this, we introduce the Malicious Discriminative Score (MDS). \textit{MDS evaluates the distinctiveness of segment behaviors, assigning higher scores to behaviors uniquely observed in malware time-series.} In Figure \ref{fig:mds}.c, the effectiveness of MDS is demonstrated. The model generates a prediction, $P_{M,W_i}$, for a test segment, and MDS ($\mathfrak{D}_{W_i}$) is independently computed based on the observed segment behavior. The MDS is low for malware-labelled benign segments (first row) and high for unique malicious segments (second row). By multiplying the MDS with the model's prediction, it either amplifies or attenuates the segment-prediction, enhancing confidence for uniquely malicious behaviors and reducing it for mis-labelled benign behaviors. Effectively, this process adjusts the classifier's hyperplane, as shown in Figure \ref{fig:mds}.b, correcting false positives arising from the mis-labelled benign segments.

\begin{figure}[!htb]
    \centering
    \includegraphics[width = 0.45\textwidth]{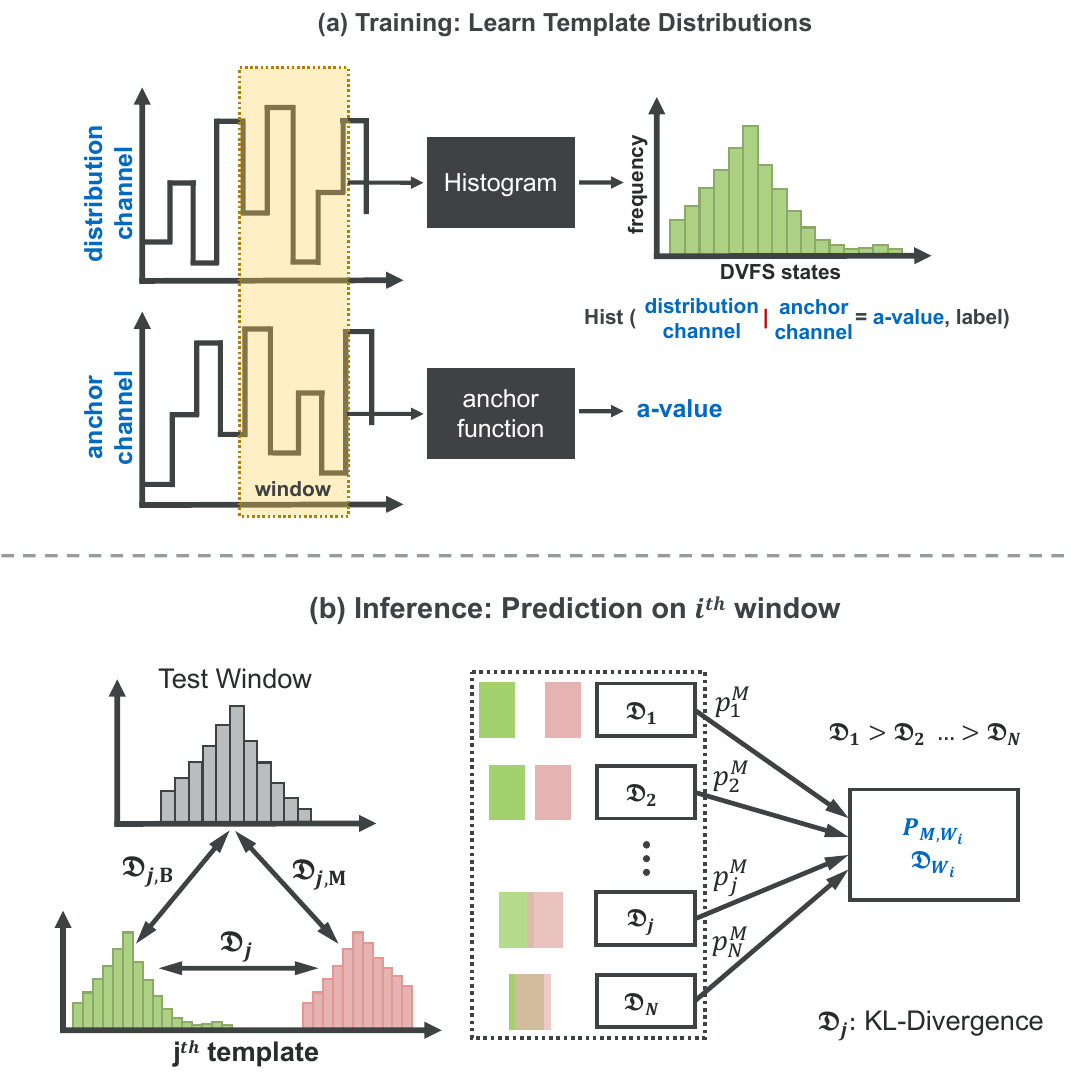}
    \caption{Illustrates the Training and Inference steps of the Interaction-based Statistical Classifier}
    \label{fig:training}
\end{figure}

\begin{algorithm}[!ht]
\caption{Training and Inference Approach}
\label{alg:training_inference}
\begin{algorithmic}[1]

\REQUIRE 
$X$: Multivariate time-series with $C$ channels \
$B$: Number of histogram bins\
$W$: Window size\
$AF$: Anchor function

\ENSURE 
$\theta$: Template distributions for training\
$P_M$: Malware probability for inference

\STATE \textbf{Training Phase}: 
\FOR{$c_1 = 1$ \TO $C-1$}
\FOR{$c_2 = c_1+1$ \TO $C$}
\STATE Select channel $c_1$ as distribution-channel
\STATE Select channel $c_2$ as anchor-channel
\FOR{non-overlapping windows $w \in W$}
\STATE $h_w \gets $ Histogram$(X_{c_1,w}, B)$ \COMMENT{Create histogram}
\STATE $a_w \gets AF(X_{c_2,w})$ \COMMENT{Determine anchor value}

\STATE $\theta_{c_1, c_2} \gets \theta_{c_1, c_2} \cup {(h_w, a_w)}$ \COMMENT{Store template}
\ENDFOR
\ENDFOR
\ENDFOR

\STATE \textbf{Inference Phase}:
\FOR{windows $w \in W$}
\STATE $\hat{h}_w, \hat{a}_w \gets$ Histogram \& Anchor$(X_{w})$ \COMMENT{Generate for distr. and anchor-channels}
\FOR{$c_1 = 1$ \TO $C-1$}
\FOR{$c_2 = c_1+1$ \TO $C$}
\STATE $\Phi_{w,B}^{c_1,c_2} \gets KL(\hat{h}_w, \theta_{c_1,c_2}^{B})$ \COMMENT{Benign KL divergence}
\STATE $\Phi_{w,M}^{c_1,c_2} \gets KL(\hat{h}_w, \theta_{c_1,c_2}^{M})$ \COMMENT{Malware KL divergence}
\STATE $P^{c_1,c_2}_{M,w} \gets \text{softmax}({\Phi_{w,B}^{c_1,c_2}, \Phi_{w,M}^{c_1,c_2}})$ \COMMENT{Malware probability}

\ENDFOR
\ENDFOR
\FOR{$(c_1, c_2) \in C\times C$}
\STATE $\mathfrak{D}_w^{c_1,c_2} \gets KL(\theta_{c_1,c_2}^{B}, \theta_{c_1,c_2}^{M})$ \COMMENT{MDS}
\ENDFOR
\STATE $P_{M,w} \gets \sum_{c_1,c_2} P_{M,w}^{c_1,c_2} \mathfrak{D}_w^{c_1,c_2} / \sum_{c_1,c_2} \mathfrak{D}_w^{c_1,c_2}$ 
\STATE $\mathfrak{D}_w \gets \sum_{c_1,c_2} \mathfrak{D}_w^{c_1,c_2}$
\ENDFOR
\end{algorithmic}
\end{algorithm}

\subsection{Training: Learning Template Distributions}
\noindent The training approach, illustrated in Figure \ref{fig:training}.a and summarized in Algorithm-\ref{alg:training_inference} Training Phase, focuses on creating template conditional distributions using empirical histograms. This process involves moving a non-overlapping sliding window across the multi-variate time-series. Within each window, a pair of channels is selected: one as the distribution-channel and the other as the anchor-channel. For the distribution-channel, a histogram based on the observed DVFS states is constructed, and for the anchor-channel, a representative anchor-value (a-value) is calculated using a predefined anchor function. Combined with the labels that distinguish between benign and malicious data sources, this forms a \textit{conditional distribution}, expressed as:

\begin{equation}
\text{P}_{\text{template}}\left[\substack{\text{distribution} \\ \text{channel}} \bigg| \left(\substack{\text{anchor} \\ \text{channel}} = \text{a-value}, \text{label}\right)\right]
\end{equation}

\textbf{Generating the Template Conditional Distributions. } As the sliding window traverses each segment of the time-series, it systematically captures key parameters of the conditional distribution such as the \textit{distribution-channel, anchor-channel, anchor-value, and label}, which are used as unique keys to retrieve and update the corresponding template histogram with new values (using add operation). This method, applied across all multivariate time-series in the training dataset creates a set of template conditional distributions, through which we have learned the likelihood of observing distributions in one channel, conditioned on a representative value in another channel given the application class (malware or benign). \textit{The MDS is defined as the Kullback-Leibler (KL) Divergence between template conditional distributions sharing the same distribution-channel, anchor-channel, and anchor-value, but differing in labels: one benign, one malicious.} Intuitively, MDS measures the uniqueness of interactions, with higher values signaling exclusive malware behavior and lower ones indicating segment-behavior common to both benign and malware time-series.

There are three key parameters used in generating the conditional distributions: the number of histogram bins (\(B\)), the window size (\(W\)), and the anchor function (\(AF\)). A grid search was performed over \(B\) and \(W\) to identify the optimal configuration which maximizes the MDS recorded over all the template distributions. The \(AF\) is used to calculate the anchor-value which is bucketized to the nearest bin to simplify the tracking of template histograms. \textit{Number of tracked template histograms: } Given \(N\) channels in a multivariate time-series from an application sample, there are \(N\) possible distribution-channel values and \(N-1\) anchor-channel values. Each anchor-value can assume \(B\) values, and distributions are tracked separately for benign and malicious labels. Thus, the overall number of template distributions monitored during training is \(N * N-1 * B * 2\).

\subsection{Inference Methodology}
The inference approach is illustrated in Figure \ref{fig:training}.b and summarized in the Inference Phase of Algorithm \ref{alg:training_inference}. It mirrors the training phase by generating inference conditional distribution for each pairwise channel. The recorded distribution-channel, anchor-channel, anchor-value parameters are used to fetch the corresponding benign and malware template distributions ($j^{th}$). The objective is to determine if the inference distribution resembles malware or benign behavior recorded by the template distributions learned during training. This is determined using the following KL-divergence scores: (1) KL divergence between the benign template distribution and the inference distribution (\(\Phi_{j,B}\)), assessing similarity to the benign behavior; (2) KL divergence between the malware template distribution and the inference distribution (\(\Phi_{j,M}\)), evaluating similarity to the malware behavior. A softmax operation, as shown in Equation \ref{eqn:softmax_malware}, calculates the malware probability:
\begin{equation}
\label{eqn:softmax_malware}
    P_{j,M} = \frac{e^{-\Phi_{j,M}}} {e^{-\Phi_{j,M}} + e^{-\Phi_{j,B}}}
\end{equation}

\noindent This operation offers a probabilistic measure of an interaction being malicious. For each pairwise interaction within the inference window (\(N \times N-1\) possible interactions), a softmax probability is calculated, reflecting the interaction’s similarity to malware behavior. The final prediction probability for the i-th window, \(P_{M,W_i}\), is obtained by averaging \(P_{j,M}\) for all \(j\) in \(N \times (N-1)\). For a uniquely malicious segment, $\Phi_{j,M} > \Phi_{j,B}$, so that \(P_{j,M} > 0.5\). However, if benign and malware interactions are similar, due to ambiguous labeling \cite{map, ensemblehmd,basu_tsc_malware_detection, low_barrier_hmd}, \(\Phi_{j,B}\approx\Phi_{j,M}\) , resulting in \(P_{j,M}\) and consequently \(P_{M,W_i}\approx0.5\). This ambiguity indicates classifier confusion due to conflicting labels, resulting in erroneous predictions.

\textbf{MDS-weighted decisions. }To address this, we use the MDS to refine our prediction process. Rather than a simple average, \(P_{M,W_i}\) is computed as a weighted average using the MDS values (\(\mathfrak{D}_{j}\)) of the corresponding template distributions, as shown in Equation \ref{eqn:mds_weighted}. This modification ensures that if benign and malware interactions are similar, resulting in a lower \(\mathfrak{D}_{j}\), the prediction probability is attenuated so that \(P_{j,M} << 0.5\). The total MDS for the i-th window (\(\mathfrak{D}_{W_i}\)) is the sum of MDS values for all participating conditional distributions, quantifying the overall MDS within the window, as shown in Equation \ref{eqn:overall_mds}.
\begin{equation}
\label{eqn:mds_weighted}
    P_{M,W_i} = \frac{\sum_{j=1}^{N \times (N-1)} \mathfrak{D}_{j} \times P_{j,M}}{\sum_{j=1}^{N \times (N-1)} \mathfrak{D}_{j}}
\end{equation}
\begin{equation}
\label{eqn:overall_mds}
    \mathfrak{D}_{W_i} = \sum_{i=1}^{N \times (N-1)} \mathfrak{D}_{j}
\end{equation}
\noindent This approach allows \(P_{M,W_i}\) to factor in the similarity of benign and malware interactions, thus refining the quality of the final prediction for each window.

\begin{figure*}[!htbp]
    \centering
    \includegraphics[width = 0.95\textwidth]{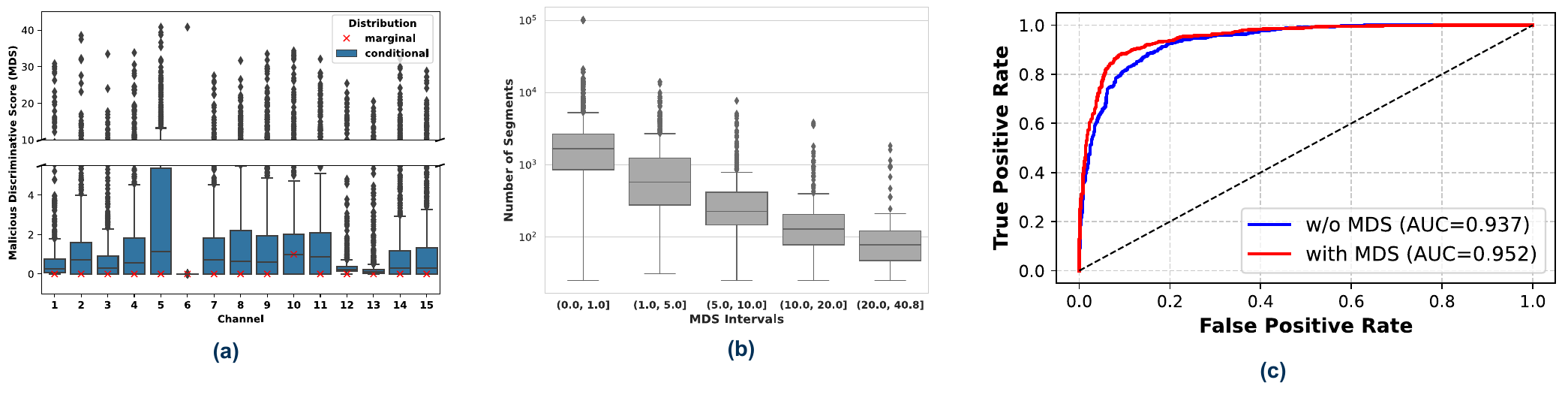}
    \caption{(a) Boxplot showing MDS of the template distributions for two modalities of information: interaction (conditional distributions) and channels of time-series treated independently (marginal distribution) for different distribution-channels. Marginals do not capture the high MDS behaviors between benign and malware; (b) Boxplot showing the frequency of occurrence of segments (total segments: 101863, segment-length$\approx$1s) stratified by MDS in the training dataset (y-axis is logarithmic). High MDS interactions are less frequent; (c) Increase in AUC upon using MDS weighted decision.}
    \label{fig:kl_divergence}
\end{figure*}

\section{Evaluation results}
\label{sec:eval_results}

\subsection{Analyzing Segment Behavior}
\textbf{Distinguishability of Interactions.}
We examine distinguishability of multi-variate interactions within a time segment, assessed via the MDS of the template distributions. 
Figure \ref{fig:kl_divergence}.a shows the MDS obtained from template conditional distributions. We observe a wide variance in MDS values indicating that the template distributions are learning both unique and common behaviors between the malware and benign time-series segments.
Channels with limited DVFS states, such as channel-5, yield lower MDS due to insufficient information for capturing interactions. The marginal template distributions which treats the channels independently as shown in Equation \ref{eqn:marginal}, are depicted with markers and consistently show lower MDS values. This underlines the necessity of leveraging conditional distributions to extract a richer set of discriminative features, highlighting the importance of inter-channel interactions over single channel analysis in a multi-variate time-series. \begin{equation}
\label{eqn:marginal}
\text{P}_{\text{template}}\left[\substack{\text{distribution} \\ \text{channel}} \bigg| \text{label}\right]
\end{equation}

\textbf{Frequency of High MDS Segments in time-series.}
Next, we aim to quantify the occurrence of high MDS segments, which are indicative of distinct malware interactions, against the backdrop of our hypothesis that a malware time-series comprises both benign and malicious segments. Figure \ref{fig:kl_divergence}.b shows the number of segments (in the training dataset) that contribute to template distributions across various MDS intervals. The y-axis, on a logarithmic scale, represents the count of segments, while the x-axis shows MDS intervals. The plot reveals a predominance of lower MDS segments, suggesting that common interactions arising from malware-labelled benign segments are frequently encountered, which substantiates our hypothesis. Conversely, segments with high MDS that reflect unique behaviors specific to malware, are rarer. This observation confirms our intuition that labeling segments solely based on the time-series they belong to—as current methods do—results in mislabeling, given the frequent commonality between segments observed in benign and malware labeled time-series. 

\subsection{Impact on Classification Performance}

\begin{table}[!htbp]
\caption{Precision and Recall for different anchor functions}
\label{tab:mds_precision_recall}
\resizebox{0.45\textwidth}{!}{%
\begin{tabular}{ccccc}
\hline
\textbf{Anchor Function} & \multicolumn{2}{c}{\textbf{Precision}}                     & \multicolumn{2}{c}{\textbf{Recall}}                        \\ \cline{2-5} 
                         & \multicolumn{1}{r}{w/o MDS} & \multicolumn{1}{r}{with MDS} & \multicolumn{1}{r}{w/o MDS} & \multicolumn{1}{r}{with MDS} \\ \hline
max    & 0.86 & 0.89 & 0.89 & 0.89 \\
mean   & 0.80 & 0.85 & 0.85 & 0.89 \\
median & 0.84 & 0.88 & 0.88 & 0.90 \\
min    & 0.83 & 0.87 & 0.89 & 0.89 \\
mode   & 0.85 & 0.90 & 0.89 & 0.89 \\ \hline
\end{tabular}%
}
\end{table}
\textbf{Classifier Performance with MDS Weighting.} Our evaluation investigates how MDS weighting influences classifier metrics, such as precision and recall, compared to baseline methods that average predictions without weighting. Due to the absence of ground truth for individual segments, we assess performance based on the entire time-series. The overall time-series prediction with MDS weighting is computed as:
\begin{equation}
\label{eqn:overall_prediction}
P_{M} = \frac{\sum_{i=1}^{S} \mathfrak{D}_{W_{i}} \times P_{M,W_{i}}}{\sum_{i=1}^{S} \mathfrak{D}_{W_{i}}}
\end{equation}

\noindent We compare the MDS-weighted decision to a baseline that aggregates predictions without MDS weighting, reflecting current practices where benign segments in malware time-series are inaccurately labeled \cite{map, ensemblehmd,basu_tsc_malware_detection, low_barrier_hmd}. We report our findings in Figure \ref{fig:kl_divergence}.c and Table \ref{tab:mds_precision_recall}. The ROC curve in Figure \ref{fig:kl_divergence}.c, with the mode as the anchor function, shows an improvement in AUC score when incorporating MDS weighted decision. Notably, for a given True Positive Rate (TPR), the MDS-weighted classifier demonstrates a reduced False Positive Rate (FPR), signifying a refined specificity. Table \ref{tab:mds_precision_recall} further reveals that across different anchor functions, MDS weighting consistently enhances precision without compromising—and occasionally improving—recall. This performance improvement aligns with our hypothesis that MDS weighting mitigates the impact of incorrectly labelled benign segments, thereby improving precision. It's important to note that while Kumar et al.'s dataset is balanced, in real-world scenarios—where benign instances vastly outnumber malicious ones—the observed precision improvement serves as a conservative estimate, as precision is particularly sensitive to the prevalence of benign samples which have a higher base rate in malware detection \cite{ml_pitfall_arp}.

\begin{figure}[!tbp]
    \centering
    \includegraphics[width = 0.49\textwidth]{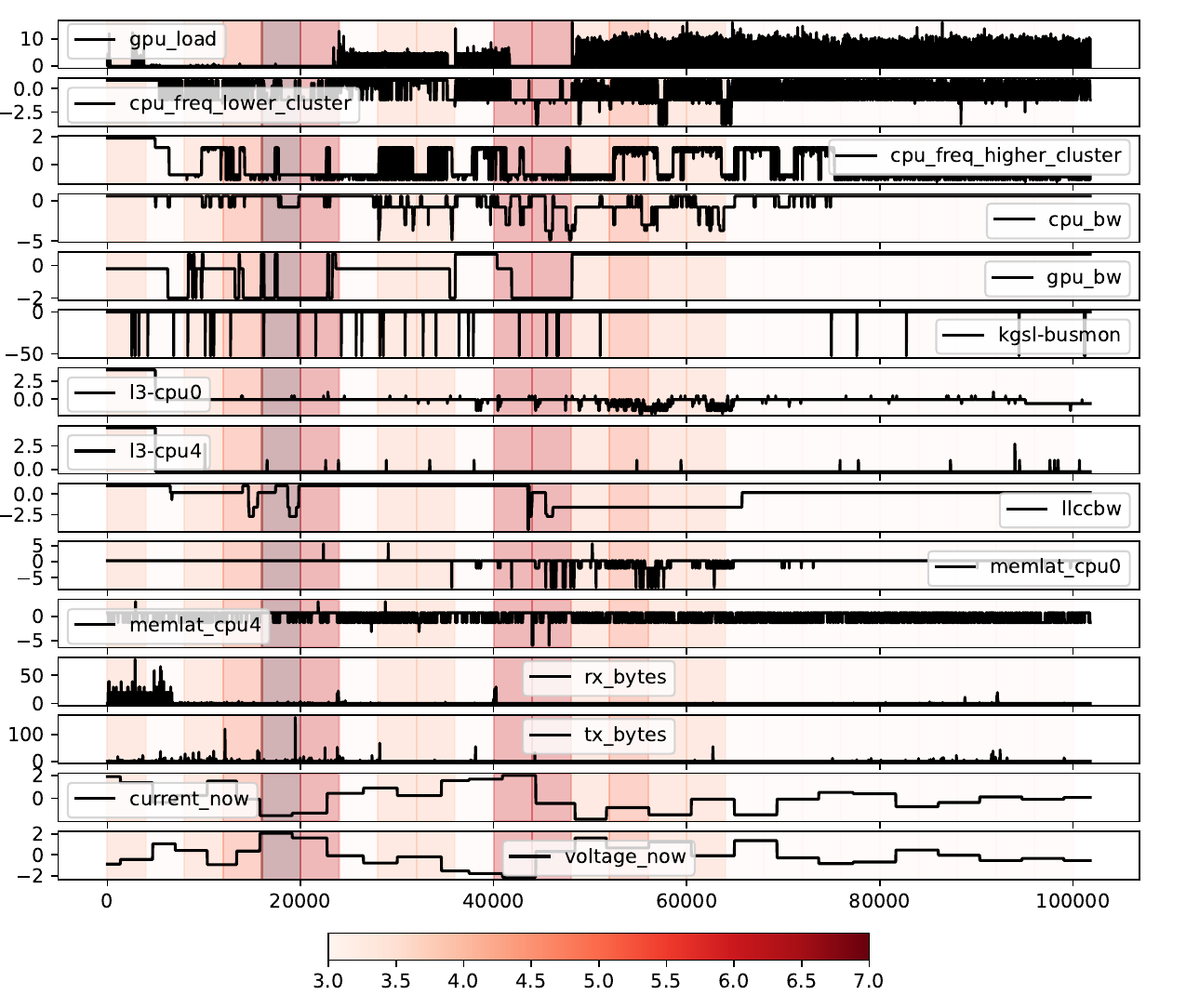}
    \caption{An MDS-based saliency heatmap highlighting localized malicious behavior within the hardware telemetry (multivariate time-series). Each channel of the time-series corresponds to the channels outlined in Table \ref{tab:dvfs_sysfs_channels}.}
    \label{fig:saliency}
\end{figure}

\textbf{Enhanced Interpretability through MDS.}
Our model introduces a refined layer of interpretability by utilizing the product of $\mathfrak{D}_{W_i}$ and $P_{M,W_i}$ as attention weights, as depicted in Figure \ref{fig:saliency}. Unlike traditional interpretability mechanisms that generate saliency maps without explaining the underlying rationale, the MDS-based saliency map assigns a \textit{definitive} meaning to these weights, suggesting that the classifier detects behavior with a high discriminative capability. 
The saliency map in Figure \ref{fig:saliency} illustrates the classifier's capability to localize malicious segments, supporting our hypothesis that not all segments of a malware time-series demonstrate distinctively malicious behavior.

\section{Conclusion and Future Work}
\label{sec:conclusion}
In conclusion, our research addresses a critical gap in HMDs by introducing a MIL framework that captures the episodic nature of malware within segmented time-series data, mitigating the label misrepresentation and subsequent false positives. Leveraging the MDS, we enhance HMD performance by distinguishing between benign and malicious segment behaviors, enhancing the precision and interpretability of threat detection. Our empirical results, demonstrated using a novel statistical classifier, confirm the efficacy of the MIL formulation in the development of robust real-time malware detection solutions. While this study concentrates on statistical classification methods applied to DVFS-based time-series, future works aim to extend MDS computation to ML algorithms and examine its viability in Hardware Performance Counter contexts, broadening the scope and applicability of the proposed solution.



\bibliographystyle{ACM-Reference-Format}
\bibliography{acmart-primary/ref}

\appendix

\end{document}